# Electron counting capacitance standard and quantum metrology triangle experiments at PTB

H Scherer, J Schurr and F J Ahlers

Physikalisch-Technische Bundesanstalt (PTB), Bundesallee 100, 38116 Braunschweig, Germany

E-mail: hansjoerg.scherer@ptb.de

**Abstract**
This paper summarizes the final results of the electron counting capacitance standard experiment at the Physikalisch-Technische Bundesanstalt (PTB) achieved since the publication of a preliminary result in 2012. All systematic uncertainty contributions were experimentally quantified and are discussed. Frequency-dependent measurements on the 1 pF cryogenic capacitor were performed using a high-precision transformer-based capacitance bridge with a relative uncertainty of 0.03 µF F$^{-1}$. The results revealed a crucial problem related to the capacitor, which hampered realizing the quantum metrology triangle with an accuracy corresponding to a combined total uncertainty of better than a few parts per million and eventually caused the discontinuation of the experiment at PTB. This paper provides a conclusion on the implications for future quantum metrology triangle experiments from the latest CODATA adjustment of fundamental constants, and summarizes perspectives and outlooks on future quantum metrology triangle experiments based on topical developments in small-current metrology.

## Introduction

Modern electrical metrology exploits the universality and paramount reproducibility of Josephson and quantum Hall effects for the reproduction of electrical units [1]. Both effects, as well as single-electron tunnelling or transport (SET) effect [2, 3], provide phenomenological relations between electrical quantities such as voltage $U$, current $I$, resistance $R$, and other parameters that are known with highest precision, like integer quantum numbers and frequency $f$. The corresponding relations are $U \propto f/K_J$, $R = U/I \propto R_K$ and $I \propto Q_S f$ for the Josephson, the quantum Hall and the SET effects, respectively, with $K_J$ and $R_K$ being the Josephson and the von Klitzing constants, and $Q_s$ the charge quantum transported in SET circuits. According to the rationale given in [4], these experimentally determined *phenomenological* constants are related to the *fundamental* constants elementary charge $e$ and Planck constant $h$ by physical theories which predict $K_J = 2e/h$, $R_K = h/e^2$ and $Q_S = e$. Although, at present, no theoretical arguments are known that predict deviations from these strict identities, the possibility of corrections must in principle not be neglected and should be tested experimentally. This is of particular metrological importance regarding the future international system of units (SI). After its impending revision, the SI will be based entirely on linking the definitions and the realization of all units to defining constants and quantum effects, respectively. These defining constants are fundamental constants like $e$ and $h$ and other constants of nature that are considered to be invariant [5].

Experimental means to test the consistency of the three quantum electrical effects are provided by quantum metrology triangle (QMT) experiments, as first proposed around 1985 [6], together with the advent of SET circuits [2]. Since then, different variants of experimental QMT realizations were and are pursued at several National Metrology Institutes (NMIs) worldwide [4, 7–14]. Typical for all QMT experiments is that the consistency of the quantum electrical effects is tested by checking the equality $K_J R_K Q_S = 2$ for the product of the phenomenological constants [15].

The first and still most successful QMT experiment was realized via the electron counting capacitance standard (ECCS) pioneered by the National Institute of Standards and Technology (NIST, USA) [7, 8, 11]. The principle of the ECCS is to charge a capacitor with a known number of electrons and to measure the resulting voltage across the capacitor electrodes. Charging of the capacitor is performed by using an SET pump which transfers charge quanta one-by-one between the capacitor electrodes. Both the SET pump and the capacitor are cryogenic elements and operated at mK temperatures in a dilution refrigerator. With the voltage being measured traced to the Josephson voltage standard (JVS) and the capacitance being measured in terms of its impedance traced to the quantum Hall resistance (QHR), the ECCS links all three quantum electrical effects and, thus, provides a QMT realization. In 2007, the final evaluation of the uncertainty budget of the ECCS-1 experiment at NIST yielded a standard uncertainty of about nine parts in $10^7$, which is also the best result so far for 'closing' the QMT [11].

By that time, the Physikalisch-Technische Bundesanstalt (PTB, Germany) was setting up a similar, but in detail different, ECCS experiment, with the aim of closing the QMT at lower uncertainty [9, 16–18]. The two main differences between the NIST and PTB setups were the metallic SET pumps (based on tunnel junctions) and the cryogenic capacitors used. The SET pump with five tunnel junctions used in the PTB experiment was equipped with on-chip micro-strip resistors in series with the tunnel junctions, following a concept known as *R*-pump [17, 19]. The resistors suppress parasitic co-tunnelling, i.e. error events. In comparison to the seven-junction pump (without resistors) used in the NIST experiment, using a pump with a smaller number of junctions was considered a practical advantage in the tuning procedures necessary for operating multi-junction pumps: a pump with *j* tunnel junctions comprises ($j$ - 1) charge islands, so the voltages on ($j$ - 1) servo gate electrodes need to be tuned for adjusting the pump's working point. The cryogenic 1 pF vacuum gap capacitor (cryocap) used in the PTB experiment had a coaxial design [20] in contrast to the NIST experiment, which used a parallel-plate-like electrode arrangement [21]. Also, the electrodes of the PTB cryocap were made from stainless steel instead of copper, and their distance was 100 times larger than in the NIST design (5 mm instead of 50 μm). A conservative estimate based on a model [22] had shown that this significantly reduces the effect of surface contaminations on the frequency dependence of the capacitance [16].

In 2011, the first successful runs of the ECCS experiment at PTB were completed. They gave results with a standard uncertainty of about 2 parts in $10^6$ and closed the QMT with the same uncertainty [13]. Being about two times higher in uncertainty than the previous best QMT result from NIST, the first PTB result, however, was considered preliminary because some contributions to the uncertainty budget were based on very conservative estimates. Also, the experimental procedures had still to be optimized, and some technical features of the PTB experiment had not yet been fully exploited. In the following period, PTB aimed at further improving the accuracy of the ECCS. The target was an uncertainty of 5 parts in $10^7$ or better. This benchmark figure corresponded to the uncertainty of the possible correction factor for $K_J$, derived from the CODATA analysis from 2010 [15, 23].

This paper summarizes the latest results on the ECCS, achieved at PTB since the preliminary result published in [13]. Improvements are highlighted, and systematic uncertainty contributions were experimentally quantified as discussed in the following. In particular, a crucial problem related to the cryocap was found, which turned out to be the show-stopper for the experiment and eventually caused the discontinuation of the ECCS at PTB. Finally, based on results from the latest CODATA adjustment [24], the paper summarizes the implications for future QMT experiments and gives an outlook on possible future developments based on recent advances by PTB in the field of high-accuracy current measurements on SET pumps.

## ECCS principle and experimental setup

The ECCS principle is based on the setup shown in figure 1. For details on setup and experimental procedures see [8, 11, 13, 16, 25] and references therein. The experiment is performed in two phases: in the first phase, the electron counting phase, the cryocap is cyclically charged/discharged by using the SET circuit. In each cycle, with the needle switch NS1 closed, the SET pump transfers *N* electrons (each with charge $Q_s$) between the capacitor electrodes. An SET electrometer is capacitively coupled on-chip to the pump via the 'pad' electrode. It monitors the voltage across the pump and drives a voltage feedback circuit (a controller with an integrator stage to eliminate dc offset errors) connected to the 'high potential' electrode of the cryocap. Thus, the electrometer serves as a null detector, keeping the voltage drop across the pump negligible to maintain a necessary condition for correct function of the pump. Also, this ensures that all transferred charge is collected on the capacitor electrodes, and not on the stray capacitances. In this way, the cryo-capacitance is calibrated by using the SET circuit and a voltmeter according to

$$C_{cryo} = C_{ECCS} \equiv NQ_S/\Delta U_{cryo} \quad (1)$$

The voltage change $\Delta U_{cryo}$ corresponding to the charging state of the cryocap is measured in terms of a JVS, providing the link to $K_J$.

In the second phase, the bridge phase, needle switch NS1 is opened and NS2 is closed to establish a connection to the 'low potential' electrode of the cryocap. In this state, the capacitance of the cryocap can be measured with a capacitance bridge: $C_{cryo} = C_{bridge}$. The impedance of the cryocap is calibrated in terms of the QHR, which provides the link to $R_K$. Comparing both independent calibrations, the QMT is 'closed' via evaluating the relation $C_{bridge}(R_K) = C_{ECCS}(Q_s, K_J)$.

**Figure 1.** Setup scheme of the ECCS. The chip with the SET circuit (comprising SET pump and electrometer) and the cryocap (1 pF) are operated in a dilution refrigerator at a base temperature of about 30 mK. Different configurations of the mechanical needle switches NS1 and NS2 correspond to the measurements for capacitor charging ($C_{ECCS}$, solid arrows) and to the measurement of $C_{bridge}$ with a capacitance bridge (dashed arrows), respectively. For preliminary tuning and performance characterization of the SET circuit, both switches are opened. Additional bias ($\pm U_{EM}/2$) and servo gate ($V_{EM}$) circuitry, connected to the electrometer via the terminals with dashed circles, is not shown in the figure for simplicity. For further details see text.

## Measurements and results

Here, results are presented from measurements that were performed on the ECCS at PTB since the publication of [13], together with experimental results from a $C_{ECCS}$ measurement campaign that extended over a period of about 2 weeks. Also, several systematic (type B) uncertainty contributions regarding the cryocap and the voltage measurement were exper imentally determined, presented in the following and discussed in more detail in the appendices F and G. Furthermore, the results from $C_{bridge}$ measurements taken in several measurement campaigns and using different capacitance bridges and techniques are discussed. Their analysis revealed a problem related to the cryocap, which hampered the successful continuation of the ECCS towards the target uncertainty crucially and finally gave rise to the abandonment of the experiment.

### 3.1. Electron counting phase

The electronic equipment (pump drive electronics and voltage feedback circuit) used for capacitor charging with the SET pump is explained in detail in [25–27]. The cryogenic setup including rf filtering is described in [13] and references therein. Experiments were performed at the base temperature of the dilution refrigerator, which is about 30 mK.

*3.1.1. Transfer error and hold time of the R-pump*

Preliminary to the capacitor charging (i.e. the $C_{ECCS}$ measurement) phase, the working point of the *R*-pump was tuned in order to achieve proper single-electron pumping and to minimize the rate of pump errors. Also, the 'hold' performance of the SET pump is important: between each two periodically repeated capacitor charging cycles, the pump is stopped and the voltage across the cryocap electrodes is measured. The single-electron dwell times on the electrodes must be sufficiently long in order to avoid charge draining from the capacitor electrodes. Following the method described in [25, 26], the tuning procedure is performed by trimming the dc voltage offsets on the four gate electrodes of the five-junction *R*-pump while monitoring the charge state of the pad (the node between the pump and the electrometer) with the SET electrometer [13]. For further details see appendix A.

Figure 2 shows results from a series of relative transfer error (RTE) measurements during the measurement campaign, derived from single-electron error counting in shuttle pumping mode, i.e. by cyclically pumping one electron onto/ from the pad island while monitoring the feedback voltage applied to the electrometer servo gate at a shuttling frequency $f_{shuttle}$ = 2.5 MHz. Results from measurements taken just before ECCS capacitor charging (i.e. right after pump tuning) and directly after each ECCS show reasonable agreement, which shows that the pump working point is sufficiently stable during the duration of an ECCS run (maximum of about 1 h). Typical RTE values of few parts in $10^8$ were found, which is of the same order of magnitude as NIST had obtained with the seven-junction pump [11, 28]. In comparison to the results published in [13], this is an improvement of about one order of magnitude. Also, note that compared to [13], these lower RTE figures were obtained with the same SET device, however, the RTE measurements shown in figure 2 were performed at five times higher shuttling frequency. In the ECCS uncertainty budget (section 3.4), a relative standard uncertainty of $1 \times 10^{-7}$ was entered for the contribution of single-electron transfer errors. Further aspects on pumping error quantification are discussed in appendix B, where also an advanced method for the estimation of the ECCS uncertainty caused by pump errors is discussed.

Average hold times, determined as part of the pump tuning processes performed during the measurement campaign, typically were of the order of several minutes (see figure A4 in appendix A) [29]. This was sufficiently long for the following ECCS capacitor charging experiments, and about 10 times better than the earlier PTB values reported in [13].

Altogether, the improvements regarding RTE figures and hold times for the five-junction *R*-pump are mainly attributed to two changes implemented after the publication of [13]. Firstly, a more elaborate iterative procedure for the pump gate tuning process was applied, which reliably enabled finding more stable working points.

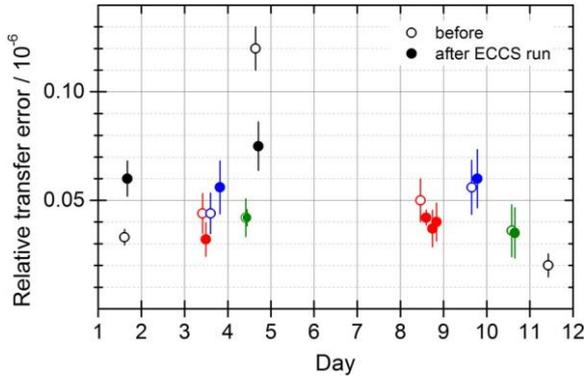

**Figure 2.** Results from measurements of the pump's RTE, determined in shuttle pumping mode at $f_{\text{shuttle}}$ = 2.5 MHz. Open (filled) symbols show values measured before (after) each ECCS run (ECCS runs are colour-coded). Error bars correspond to the statistical uncertainties, see figure A2 in appendix A.

Secondly, the sample box in the dilution refrigerator (i.e. the cold box mounted to the mixing chamber of the cryostat, containing SET chip, cryocap and needle switches) was wrapped in several layers of aluminium foil to provide additional screening of the SET device from rf interference, for instance from thermal radiation that could leak through small slits in coaxial connectors on the sample box. Parasitic tunnelling and background charge activity in the SET device, typically caused by electromagnetic or thermal activation, were possibly reduced by this. In the literature, however, different and sometimes contradicting results regarding rf screening in SET experiments are reported: in a study on the seven-junction SET pump used at NIST [30], the error rate was found unchanged if the rf shield around the SET circuit was completely removed. It was thus concluded that the source of high-frequency photons triggering the errors was inside the screening box. In a more recent study on a single-electron trap device, however, it was shown that the hold times could be increased drastically by heavy screening of the SET circuit, technically provided by two nested rf-tight radiation shields [31]. This result was explained by the suppression of photon-assisted tunnelling events, triggered by high-frequency photons from outside the sample box.

*3.1.2. Voltage measurement setup*
For the measurement of $U_{\text{cryo}}$ (see figure 1), a commercial 8½-digit voltmeter (type Agilent model 3458A) was used. During the ECCS measurement campaign discussed in the following section 3.1.3, the voltmeter gain was calibrated over the range ±10.5 V typically twice per day (i.e. before and after the ECCS runs) using a programmable JVS system. Over the measurement campaign period of 11 d, the voltmeter showed a gain stability corresponding to a relative scatter of less than 0.2 µV V$^{-1}$. For the evaluation of the ECCS data, gain values from calibrations taken before and after the ECCS runs were interpolated with uncertainties < 0.04 µV V$^{-1}$. Conservatively, for the traceability of $U_{\text{cryo}}$ to $K_J$, a standard uncertainty of 0.05 µV V$^{-1}$ was assigned in the uncertainty budget (section 3.4).

After the ECCS campaign presented in this paper (see section 3.1.3), the voltage measurement setup was improved by introducing a system similar to one presented in [32]. The programmable JVS system in combination with the voltmeter was configured to enable $U_{\text{cryo}}$ measurements differentially: the JVS is used to compensate $U_{\text{cryo}}$, and the remaining small voltage difference is measured with the voltmeter in its lowest (100 mV) range. In this way, the requirements on voltmeter gain calibration and stability are relaxed significantly, and $U_{\text{cryo}}$ measurements at 10 V level could be performed with a total relative uncertainty < 1 nV V$^{-1}$ [29].

*3.1.3. Capacitor charging by single-electron counting.*
Following the pump tuning procedure (see section 3.1.1 and appendix A) providing high-fidelity performance of single-electron transfer, sufficiently long hold times and stability of the working point, the capacitor charging phase of the ECCS was initialized. For this, needle switch NS1 was closed. Next, the voltage feedback for $U_{\text{cryo}}$ (see figure 1) was activated. This step is crucial since it has to be done without disturbing the voltage $U_{\text{pump}} \approx 0$ across the pump to maintain proper pumping performance [27]. The setup procedure for the feedback followed the steps described in [11]: first, $U_{\text{pump}} \approx 0$ was established by tuning the working point of the SET electrometer (in this phase serving as a null detector for $U_{\text{pump}}$) by trimming the voltage $V_{\text{EM}}$ on its servo gate ($C_{\text{servo}}$, figures 1 and A1) to minimize the error signal of the feedback circuit. This avoided abrupt changes of $U_{\text{cryo}}$ when the feedback was locked. During locking, $U_{\text{cryo}}$ was monitored to check if the voltage across the pump did not exceed the critical $U_{\text{pump}}$ value of about 10 µV (see appendix C).

The relation between both voltages, determined by capacitive coupling in the circuit, was $U_{\text{cryo}} \approx 10\, U_{\text{pump}}$. After successful locking, the following capacitor charging generally worked well. The locking attempt was repeated if a voltage jump exceeding the critical level was observed. If this had to be repeated several times, pumping performance was checked by returning to an RTE measurement, and, when necessary, the pump was re-tuned.

The pump drive electronics was configured for a clock frequency of 2.5 MHz. The number $N$ (tens of millions) of electrons to be transferred between the cryocap electrodes was determined by corresponding settings of hex pots on the pump drive. Capacitor charging, with the feedback voltage $U_{\text{cryo}}$ being monitored continuously, was started by pumping $N/2$ electrons in one direction, followed by pumping ±$N$ electrons repeatedly for cyclic charging symmetrically around $U_{\text{cryo}} = 0$, as shown in figure 3. In between the up and down charging ramps, i.e. after $N$ electrons had been pumped, the pump was stopped for several seconds to measure $U_{\text{cryo}}(\pm N)$.

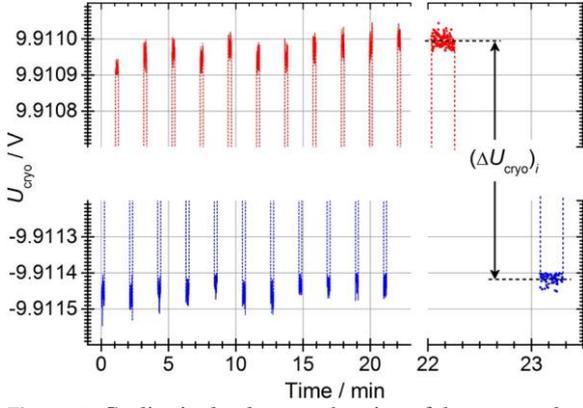

**Figure 3.** Cyclic single-electron charging of the cryocap between about ±10 V. The full trace represents one 'ECCS run'. The pump electronics was configured to transfer $N$ = 123 731 968 electrons (7600000 on pump drive hex pots) at a clock frequency of 2.5 MHz during each charging ramp, corresponding to a ramping time of about 50 s. After each ramp, the pump was stopped for 13 s for measuring the 'plateau' values $U_{cryo}(\pm N)$, as shown in the expanded view on the right-hand side of the panel. The RTE determined from measurements before and after this run was $4 \times 10^{-8}$, corresponding to an uncertainty of five electrons transferred per ramp.

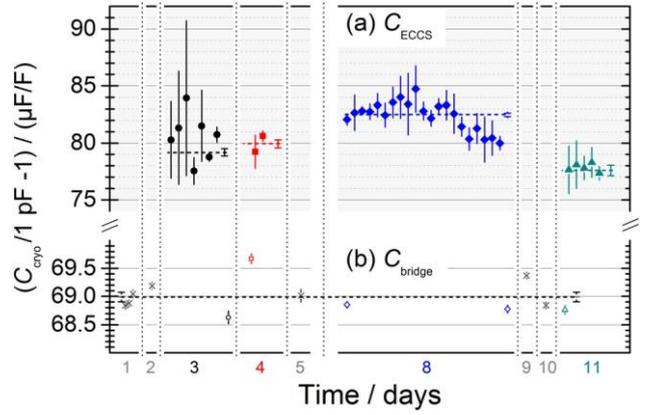

**Figure 4.** (a) $C_{ECCS}$ results of 35 runs from the ECCS, performed on days 3, 4, 8 and 11 of the measurement campaign (colour coded) without thermal cycling of the system. Data are shown in terms of the relative deviation the cryocap capacitance from 1 pF. Error bars correspond to type A standard uncertainties ($k$ = 1) including scatter and asymmetry of the voltage plateau data. The weighted means of the data for each of the 4 days are shown as dashed lines together with their uncertainties (small capped error bars right to the right of the data), which ranged between ±0.15 µF F$^{-1}$ (day 8) and ±0.46 µF F$^{-1}$ (day 11). (b) Results from 12 $C_{bridge}$ measurements performed with the AH bridge, each from data acquired and averaged over 8 min up to 53 min, with error bars corresponding to type A standard uncertainties between 0.05 µF F$^{-1}$ and 0.13 µF F$^{-1}$. Results from data taken on days without ECCS measurement activities are shown as cross symbols. The mean value (68.9 ± 0.1) µF F$^{-1}$ is indicated by the horizontal dotted line, with an uncertainty corresponding to the standard deviation of the mean (SDOM, shown as capped error bars).

During the initial phase of each capacitor charging run, the plateau values typically tended to drift to one direction, as shown in figure 9 of [27] (not visible in figure 3). This indicates asymmetric pumping caused by the pumping errors being slightly different for both directions. Such asymmetry is typically caused by a small residual voltage $U_{pump}$ across the pump [11, 27]. It was compensated by slight trimming of the voltage $V_{EM}$ applied to the electrometer servo gate, coupled to the pad node on-chip via $C_{servo}$ and $C_{ID}$. In this way, $U_{pump}$ was fine-tuned until the drift of the plateau voltages disappeared and showed non-monotonic random behaviour, as visible in figure 3. In some runs, the trimming had to be repeated. For the evaluation of the $U_{cryo}$ traces discussed in the following, the data from the trimming phases were discarded. Remaining asymmetry effects were treated according to the rationale given in [11], and a corresponding uncertainty contribution was accounted. Details on the effect of voltage bias across the five-junction $R$-pump are discussed in appendix C.

The end of an ECCS run was indicated by a sudden increase in pumping asymmetry. This is caused by fluctuations and drift of the background charges shifting the SET electrometer working point until the feedback circuit could no longer maintain $U_{pump}$ sufficiently small. Typically, and similar to the NIST experiments reported in [11], this limited the duration of ECCS runs to about 1 hour. After ending an ECCS run, the setup procedure for the feedback explained above was repeated and the next ECCS run started. Between subsequent ECCS runs, the pumping and hold performance of the pump was checked according to the procedure explained in appendix A. Re-tuning of the pump was necessary about once per day. Over a period of 9 days, 35 ECCS runs with durations between 4 min and 65 min were performed in one cooldown cycle, i.e. without thermal cycling of the system, with $U_{cryo}$ = ±5 V ($N$ = 61 865 984, hex 3B00000) and ±10 V ($N$ = 123 731 968, hex 7600000). Data from each ECCS run were processed according to the algorithms described in detail in section 4.1 of [11], based on the differences $(\Delta U_{cryo})_i$ of the mean $U_{cryo}$ plateau values from the $i$th ramp (see figure 3). Averaging these data yielded the total result for $\Delta U_{cryo}$ of each run, together with type A uncertainty contributions comprising the scatter of $(\Delta U_{cryo})_i$ and the contribution from the plateau asymmetry between up and down ramps. Finally, for each run, $C_{ECCS}$ in SI farad was derived according to equation (1), setting $Q_S = e$ and taking into account that $\Delta U_{cryo}$ was measured in terms of $V_{90}$ (the voltmeter was calibrated in terms of $K_{J-90}$) by applying the unit transformation $\{U\}_{SI} V = \{U\}_{90} V_{90}$ [13]:

$$C_{ECCS} = N (2e^2/h)/(\{\Delta U_{cryo}\}_{90} K_{J-90}) \text{ volt} \qquad (2)$$

Values for $e$ and $h$ in SI units from the latest CODATA adjustment [24] were used for the evaluation of equation (2). Figure 4(a) shows the $C_{ECCS}$ results in terms of the relative deviation from 1 pF.

The $C_{ECCS}$ results from the first five ECCS runs of the measurement campaign (first five data points from day 3) show type A uncertainties up to 7 µF F$^{-1}$ due to relatively large voltage plateau asymmetry contributions. In later runs, smaller asymmetries were achieved by improved fine-tuning of $U_{pump}$, and total type A uncertainties ranging between 0.5 μF F$^{-1}$ and 2 μF F$^{-1}$ were achieved. This is an improvement compared to the results from the four preliminary ECCS runs presented in [13], which showed type A uncertainties between 2.5 μF F$^{-1}$ to 3.9 μF F$^{-1}$. However, the results for $C_{ECCS}$ from different days exhibit scatter of few μF F$^{-1}$ (see the weighted mean values shown in figure 4) and also drift on the time scale of a day (see the data acquired during day 8). Since the parameters $N$ and $U_{cryo}$ entering $C_{ECCS}$ according to equation (2) have type B uncertainties considerably smaller than 1 μF F$^{-1}$, the results indicate problems either with the ECCS measurement procedure or with the cryocap. Further discussion of this issue is resumed in section 3.3.

### 3.2. Bridge phase

As explained in the section 1, the interpretation of ECCS results in terms of a QMT experiment requires tracing $C_{cryo}$ to $R_K$ using capacitance measurement techniques. For the experiments described in the following, different commercial ac bridges as well as a coaxial (transformer-based) capacitance bridge, designed at PTB for impedance metrology at highest accuracy, were used. The cryogenic setup for cryocap measurements in the bridge phase is described in [13] and references therein. Additional information on the cryogenic wiring and the grounding scheme is given in appendix D.

During the measurement campaign, $C_{bridge}$ was monitored with a commercial precision ac capacitance bridge (Andeen-Hagerling model 2500A, abbreviated 'AH' in the following) at a fixed frequency of 1 kHz and with an effective excitation voltage of 15 V(rms). On the days when $C_{ECCS}$ measurements were performed, $C_{bridge}$ was measured directly before and/or after the ECCS runs. The bridge was calibrated in terms of $R_{K-90}$ with an uncertainty less than 1 μF F$^{-1}$. The results were converted to SI farad and are shown in figure 4(b) in terms of the relative deviation from 1 pF. The drift derived from a linear fit over the data from 11 d was < 0.02 μF F$^{-1}$ per day (in agreement with earlier investigations using the same bridge instrument [20]), and the data scatter was within ±0.8 μF F$^{-1}$. Since the stability of the bridge is better than 1 μF F$^{-1}$ per year (corresponding to ≈ 3 nF F$^{-1}$ per day), we attribute the scatter of the $C_{bridge}$ data to fluctuations of $C_{cryo}$ itself and/or to effects from the cables between the cryocap and the bridge. This implies that $C_{bridge}$ measurements generally have to be performed at least daily, preferably shortly before and/or after $C_{ECCS}$ measurements.

### 3.3. Comparison between ECCS and bridge phases

Generally, for a quantitative comparison of $C_{ECCS}$ and $C_{bridge}$ values, several effects related to the different measurement conditions in both phases have to be considered. Firstly, in the ECCS phase the cryocap is charged at an effective frequency of the order of 10 mHz (corresponding to the period of the charging cycles), while capacitance bridges typically work with ac excitation in the kHz range. For this reason, the possible intrinsic frequency dependence of $C_{cryo}$ due to surface contamination of the cryocap electrodes has to be quantified over several decades of frequency. A conservative estimate based on a model discussed in [22] had shown that for the PTB cryocap the uncertainty due to this frequency dependence is smaller than 20 nF F$^{-1}$ [16]. Secondly, single-electron capacitor charging and bridge measurements are typically carried out at different voltage levels, and a possible voltage dependence of $C_{cryo}$ needs to be considered. Thirdly, the cryocap in the ECCS phase is charged by the SET pump through wires of few centimeters of length, while the capacitance bridge measurements involve different and segmented cables of several meters of length, partly installed inside the refrigerator and furthermore heavily rf-filtered (see appendix D). Thus, cable effects and corresponding corrections to the $C_{cryo}$ values measured with the bridge have to be accounted for.

Comparison of the results shown in figure 4 exhibits a significant discrepancy between $C_{ECCS}$ and $C_{bridge}$ of the order of about 10 μF F$^{-1}$. Such discrepancy was not found in the ECCS experiments, performed with the same setup and instruments, presented in [13]. However, subsequent investigations indicated that the geometrical arrangement and length of the coaxial cables connecting the AH bridge and the terminals on top of the refrigerator influence the $C_{bridge}$ measurement results: effects of the order of few μF F$^{-1}$ were noticed, which is of the same order as the cable correction of 1.5 μF F$^{-1}$ that was applied for measuring $C_{bridge}$ in the ECCS experiment at NIST using a bridge of the same type [11]. Note that the AH bridge does not operate with equal and opposite return currents in each outer conductor, so the cable effect is much larger than for a coaxial bridge and also depends on the spatial routing of the cables. In an attempt to shed further light on the cable effects related to the measurements with the AH bridge, measurements with a commercial multi-frequency bridge (AH model 2700A) were performed for frequencies up to 20 kHz. The results shown in appendix E verify that the cable effects of the AH bridge are of the order of 10 μF F$^{-1}$ at a frequency of 1 kHz, which could explain the apparent discrepancies between $C_{bridge}$ and $C_{ECCS}$.

Altogether, the findings implied that the limitations given by the commercial bridge impaired the $C_{bridge}$ measurements significantly, and further motivated the use of a coaxial capacitance bridge technique available at PTB which is less sensitive to cable effects and far more accurate. In addition, the inductive divider bridge enabled measurements of the frequency dependence of $C_{cryo}$. These investigations are discussed in the following.

#### 3.3.1. Measurements with the coaxial ratio bridge

The impedance bridge used for the following investigations is a high-accuracy inductive coaxial ratio bridge (ratio 1:10), used at PTB routinely as a part of the measuring chain for realizing the farad from two ac QHR resistors. The coaxial design of the ratio bridge follows the principles explained in [33] and specific details are given in [34]. It was demonstrated that this bridge is capable of tracing a capacitance of 1 pF to the QHR with a relative uncertainty of 20 nF F$^{-1}$ via comparison with a 10 pF fused-silica capacitance standard, which exceeds the accuracy of the commercial instrument by about two orders of magnitude (the AH bridge is specified with a standard uncertainty of 1.25 µF F$^{-1}$). In the ECCS uncertainty budget (section 3.4), a relative standard uncertainty of 0.03 µF F$^{-1}$ was entered for the contribution of the ratio bridge measuring $C_{cryo}$, including small additional systematic uncertainty contributions for the voltage dependencies in the measuring chain. The effects of the cryogenic filtered cabling, described in appendix D, on capacitance measurements with this bridge had been analyzed and found to be small: for $C$ = 1 pF and $f_{bridge}$ = 1233 Hz, a cable correction corresponding to $\Delta C/C$ = (20 ± 1) nF F$^{-1}$ was determined [16]. Note that this value and its uncertainty are about two orders of magnitude smaller than the cable corrections for the AH bridge (see [11]), and that it is well-defined and independent of the spatial routing of the cables.

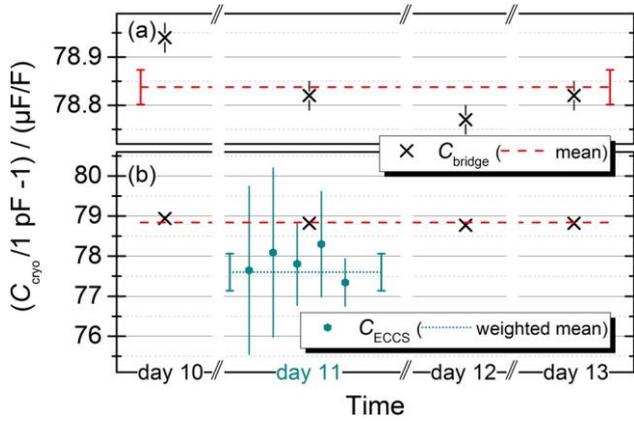
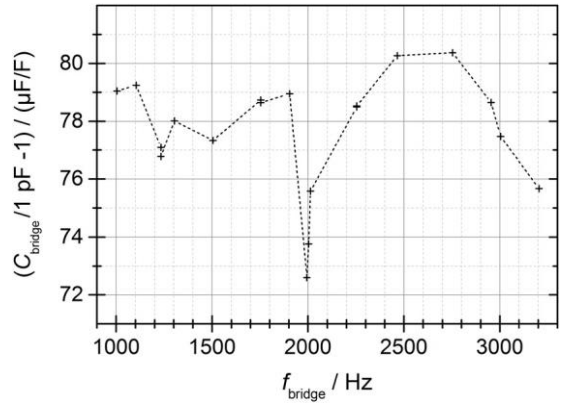

**Figure 5.** $C_{bridge}$ measured with the high-accuracy ratio bridge at $f_{bridge}$ = 1233 Hz and an effective excitation voltage of 15 V(rms) in comparison with $C_{ECCS}$, measured around the end of the ECCS measurement campaign without thermal cycling. The $C_{bridge}$ data (a) have error bars corresponding to 30 nF F$^{-1}$ combined uncertainty for each of the four bridge measurements, and their mean value, indicated by the dashed line, has a standard uncertainty of 36 nF F$^{-1}$ (SDOM, capped error bars). (b) $C_{bridge}$ together with $C_{ECCS}$ (results of day 11, also shown in figure 4(a)), and their weighted mean (dotted green line with capped error bars).

**Figure 6.** $C_{bridge}$ measured with the high-accuracy ratio bridge at different frequencies in the audio range and with an effective excitation voltage of 15 V(rms) at base temperature in the refrigerator. The data were taken in a separate cooldown cycle after the measurements shown in figures 4 and 5. Error bars corresponding to measurement uncertainties are smaller than data symbols and not visible in the graph. The dashed lines interconnecting the points are just a guide to the eye. For proper measurements at each frequency, the phase of the detector was aligned without the usual phase-shifted injection system, and the 10:1 ratio was interpolated to the particular frequency.

First $C_{cryo}$ measurements with the ratio bridge were performed over 4 days around the end of the ECCS measurement campaign, and the results are shown in figure 5. The $C_{bridge}$ data scatter of about 0.2 µF F$^{-1}$ (figure 5(a)) over 4 days is smaller than the scatter of typical results measured with the AH bridge (see figure 4(b)). This indicates that the ratio bridge is indeed the superior instrument for $C_{bridge}$ measurements. Also, compared to the results from the measurements with the AH bridge, the discrepancy between $C_{ECCS}$ and the $C_{bridge}$ results measured with the ratio bridge was smaller, but still significant: as figure 5(b) shows, a remaining difference of 1.2 µF F$^{-1}$ between the mean values of $C_{bridge}$ and $C_{ECCS}$ was found. This finding hinted at the presence of a frequency-dependent effect in the cryocap measurements which needed further investigation.

Initially designed for the operation at fixed frequencies of 1233 Hz and 2466 Hz, the ratio bridge technique was modified around 2012 to enable measurements in the frequency range between about 500 Hz and 3 kHz. Consequently, $C_{bridge}$ measurements at different frequencies $f_{bridge}$ in the range from 1 kHz up to about 3 kHz were performed. The results shown in figure 6 exhibit a complex frequency dependence with pronounced peak-like structures of the order of 10 µF F$^{-1}$ in magnitude. Regarding the pursued target of the experiments, i.e. 'closing' the QMT via the ECCS at improved accuracy, this finding is dramatic since the $C_{bridge}$ values need to be extrapolated to the dc regime of the ECCS phase with an uncertainty well below 1 µF F$^{-1}$. Therefore, clarification of the reason for this unexpected finding and a corresponding remedy were crucial for the project and pursued intensively.

First, the influence of the main elements of the setup were investigated by measurements with the ratio bridge. For this, the coaxial cabling, including the rf filters installed inside the refrigerator, was tested at room and at base temperatures in measurements on a commercial 1 pF fused-silica capacitance standard, and only a small frequency dependence as expected from the cable parameters was found. Also, the rf filters were separately tested at room temperature (RT), with the result that they did not cause the strong frequency dependence observed in the measurements shown in figure 6. The same finding resulted from separate measurements of the cryocap, performed at RT and connected directly to the ratio bridge.

Next, the needle switch NS2 (see figure 1) was removed from the cryogenic setup, and the cryocap was connected to the cryogenic cables at the mixing chamber stage of the refrigerator by a short piece of coaxial line. Following $C_{\text{bridge}}(f)$ measurements at base temperature again revealed peak-like structures similar to those observed before. Interestingly, the $C_{\text{bridge}}(f)$ characteristic became smooth (i.e. the pronounced frequency dependence disappeared) when the whole system including cryocap and cabling was warmed up to RT, which seemed to hint at the idea that the operation of the dilution refrigerator caused the problem. However, this was ruled out by following experiments in which different operation modes were tested: neither different setting of the 1 K-pot needle valve made a difference (under circumstances the 1 K-pot of dilution refrigerators may cause vibrations), nor did completely stopping the helium circulation in the refrigerator provide a remedy.

With respect to the idea that the peak-like structures observed in the $C_{\text{bridge}}(f)$ dependence might be caused by mechanical or electrical resonances, another aspect has to be noted: the ac bridge technique is in principle only sensitive for interferences that are phase-coherent with the ac excitation voltage applied by the bridge. A possible corresponding scenario to be checked was if electro-acoustic coupling between 'high' and 'low' potential cables connecting the cryocap and the bridge would cause such interference. It appeared possible that the PTFE insulation in the semi-rigid coax line segments (see appendix D) installed in the refrigerator could shrink when the system was cooled, so that the inner conductor of the 'high' potential cable might oscillate, driven by the ac excitation voltage. The corresponding vibration might be picked up by the 'low' potential line connected to the detector of the bridge. In this way, excitation and resonances in-phase with the ac bridge voltage could occur and be detected by the bridge. To clear this up, eventually the cryocap was wired to the RT terminals of the refrigerator completely by commercial flexible coax cables with braided shield (type Gore™), and without involving rf filters. These cables are also used in another refrigerator used at PTB for high-precision impedance measurements based on ac quantum Hall resistors [34] and, thus, known to be unproblematic.

High-accuracy $C_{\text{bridge}}$ data measured at discrete bridge frequencies on two subsequent days (cross and star symbols in figure 7) again show pronounced frequency dependence with various structure of magnitudes of the order of several $\mu$F F$^{-1}$ up to more than 10 $\mu$F F$^{-1}$, similar to the measurements shown in figure 6. The $C_{\text{bridge}}(f)$ characteristic measured during a continuous sweep of the bridge frequency (black line in figure 7) revealed a series of strong and sharp double-peak-like structures. The data taken in discrete and continuous measurement modes on day 2 slightly disagree in part due to the limited accuracy of few $\mu$F F$^{-1}$ for the bridge being operated in continuous mode, as explained in the caption of figure 7. Altogether, these results from $C_{\text{cryo}}(f)$ measurements performed with a reliable ac capacitance bridge technique unambiguously revealed that the problem was caused by the cryocap itself. Further details on the experiments and tests performed to track down the problem are given in appendix H, where the main steps of the diagnostic process are listed.

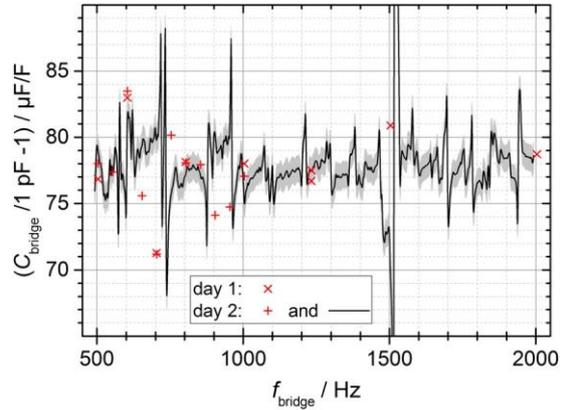

**Figure 7.** $C_{\text{bridge}}$ measured with the ratio bridge in the audio frequency range with an effective excitation voltage of 15 V(rms) at base temperature in the refrigerator after re-wiring of the cryocap with flexible coaxial cables. The red symbols show data from measurements at discrete frequencies, taken at two subsequent days. In this 'standard operation' mode, the ratio bridge was tuned for each frequency to eliminate frequency-dependent reactance components of the complex detector signal and provide highest measurement accuracy (error bars are much smaller than the symbols in the graph). The black line shows the result from measurements performed in the 'continuous' operation mode of the ratio bridge with $f_{\text{bridge}}$ being swept while the detector signal was monitored continuously. The duration of the full sweep was about 1 h. Balancing of the bridge was performed before the sweep at a frequency around the middle of the interval. Thus, the signal reactance component was not cancelled perfectly over the whole sweep range, and the measurement uncertainty was increased to about 1 $\mu$F F$^{-1}$ (grey area around the curve). The largest resonance-like structure appearing around 1.5 kHz, clipped by the figure borders, has a peak–peak amplitude of about 60 $\mu$F F$^{-1}$.

Since the structures visible in the $C_{\text{cryo}}(f)$ characteristic in figure 7 correspond to driven mechanical oscillations around their resonance frequencies, a possible explanation could be mechanical resonances of the many individual components of the cryocap, driven by the electric field. At low temperature, the individual components may not perfectly match, and also mechanical damping due to inelasticity of the materials involved is orders of magnitude smaller than at RT, which could lead to strong resonances. However, the scenario remained puzzling and the problem with the cryocap unsolved. Because no simple clues for remedy were at hand, a possible re-design of the cryocap would have required extensive further investigations with unknown outcomes. An extrapolation of $C_{\text{bridge}}$ data from the kHz frequency range down to the dc regime of the ECCS, as required for the QMT experiment, appeared impossible to perform with the required accuracy of 1 $\mu$F F$^{-1}$ or better. Therefore, the project was stopped.

**Table 1.** Type B (systematic) standard uncertainty contributions of the ECCS uncertainty budget for the PTB experiments compared to the NIST results [11]. All figures of the PTB budget are conservative estimates of upper limits.

| Type B uncertainty component | NIST ECCS-1 all values from [11] | PTB this paper | Remark |
|---|---|---|---|
| | Relative standard uncertainty ($k=1$) in parts per million | | |
| ECCS phase: | | | |
| $U_{cryo}$ traceability to $K_J$ | 0.05 | 0.05 | From section 3.1.2 |
| Cryocap charge leakage | 0.01 | 0.01 | From appendix F |
| SET pump transfer error | 0.01 | 0.1 | From section 3.1.1 |
| Bridge phase: | | | |
| $C_{cryo}$ traceability to $R_K$ (bridge accuracy only) | 0.82 | 0.03 | From section 3.3.1 |
| Loading effects | 0.3 | 0.02 | From [16] and section 3.3.1 |
| Comparison between phases: | | | |
| $C_{cryo}$ voltage dependence | 0.09 | 0.05 | From appendix G |
| $C_{cryo}$ frequency dependence due to surface contamination | 0.2 | 0.02 | From [16] and section 3.3 |
| Total type B | 0.90 | 0.13 | Root of sum of squares of the values above |
| $C_{bridge}$ extrapolation to dc (10 mHz) | (included in 'loading effects') | >1 | From section 3.3.1 |

### 3.4. Conclusions on the ECCS at PTB

Table 1 summarizes the type B uncertainty components for the ECCS experiment at PTB, all discussed in this paper, in comparison with the type B uncertainties from the NIST ECCS-1 experiment [11]. It shows that the PTB experiment was indeed competitive with the earlier NIST experiment, except for the unexpectedly large uncertainty contribution from the extrapolation of the $C_{bridge}$ results to about 10 mHz, i.e. to the effective frequency of capacitor charging in the ECCS phase. Thus, the ECCS at PTB could have closed the QMT with an uncertainty below 1 part in $10^6$ if it had not faced the cryocap resonance problem.

As about a decade of experience at PTB had shown, closing the QMT with the ECCS is a complex and demanding experimental task with multiple failure modes. By the time of invention of the ECCS principle by NIST in the early 1990s [7], SET pumps were able of generating only very small currents of the order of few pA. Thus, the principle of accumulating the charge delivered by such pump on a cryogenic capacitor to generate a voltage at the level of several volts showed an elegant way of bringing SET pumps to their metrological application, and a breakthrough idea for closing the QMT. In retrospective, it seems unfortunate that technical problems with the cryogenic vacuum capacitor finally were the showstopper for the ECCS at PTB. However, the identification of the crucial problem, i.e. the revelation of resonances in the capacitor measurements, was only possible by high-accuracy investigations of the frequency dependence of $C_{cryo}$. This was enabled by advances in the capacitance measurement technique at PTB, i.e. by the possibility of performing frequency-dependent measurements.

### Present status and perspectives of the QMT

Meanwhile, since the time the measurements presented in this paper were performed, developments have changed the QMT picture. As explained in section 1, the motivation for closing the QMT via the ECCS with an uncertainty of 5 parts in $10^7$ or better [15] was derived from the CODATA analysis from 2010 (published in 2012 [23]), which had yielded uncertainties for the possible correction factors $\varepsilon_J$ (for $K_J$) and $\varepsilon_K$ (for $R_K$) of about 5 parts in $10^7$ and 2 parts in $10^7$, respectively. In 2016, the results from the follow-up CODATA adjustments performed in 2014 were published [24]. According to this, the current uncertainties are two parts in $10^8$ for both $\varepsilon_J$ and $\varepsilon_K$. This drastic improvement was based on recent progress in the watt-balance and silicon-sphere experiments, which enabled former disagreements relevant for the determination of $\varepsilon_J$ and $\varepsilon_K$ to be resolved, as explained in [24] (p 55).

Regarding the interpretation of the most accurate QMT experiment so far, the results from the ECCS-1 experiment at NIST [11], the achieved uncertainty of about one part in a million are to be interpreted in terms of a constraint on $\varepsilon_S$, i.e. the uncertainty of knowledge on the relative difference between $e$ and the value of the charge quantum $Q_S$ transferred by the single-electron pump [35]. Regarding future QMT experiments, the latest CODATA adjustment with the new benchmark figures means that a significant impact on the Josephson and quantum Hall relations (in the sense of an increase of confidence) now requires closing the QMT with an uncertainty of one part in $10^8$ or better. Neither the former ECCS experiment developed at NIST nor at the PTB version had the potential of reaching this accuracy level. Also, at present, no concrete ideas are at hand giving a perspective towards that target. However, recent advances in SET pump technology and small-current metrology open a way for realizing an improved QMT experiment at an uncertainty level significantly below of one part in a million by measuring an SET-generated current in terms of $R_K$ and $K_J$, as described below. The result from such an experiment would be significant since it could further constrain the present uncertainty of $\varepsilon_S$.

Recent experiments at PTB have demonstrated that SET pumps based on semiconductors are capable of sourcing relatively high quantised currents of the order of 100 pA with uncertainties below 0.2 µA A$^{-1}$ [36, 37], and the error rate of these pumps is estimated to be of the order of few 1000 s$^{-1}$. Based on that, 'self-referenced' SET current standards are being developed at PTB, comprising several such pumps in series in combination with single-electron detectors [38, 39]. The term 'self-referenced' indicates that these circuits enable quantifying single-electron transfer errors during current generation—an improvement compared to the error-accounting scheme used in the ECCS, which is performed only before or after the current-sourcing phase (see section 3.1 and appendix A of this paper). As shown in [39, 40], such a circuit with three pumps in series has the potential to reduce the error rate by a factor of about 1000 compared to a single pump. At present, work at PTB is in progress to increase the detector bandwidth up to about 100 kHz by using rf SET electrometers (metallic SET transistors as well as semiconductor quantum dots and quantum point contacts). Altogether, these ongoing efforts aim at realizing a 'self-referenced' SET current standard with the potential to source a current $I_{SET} = \langle N \rangle \times Q_s \times f$ of 100 pA with a relative current uncertainty well below 0.1 µA A$^{-1}$. Here, $f$ is the nominal pumping frequency, i.e. the repetition frequency of the voltage signals driving the pump, and $\langle N \rangle$ is the average number of electron transferred per pump cycle as quantified by error-accounting using rf single-electron detectors. With an alternative *in situ* error-counting scheme developed at the National Physical Laboratory (NPL, UK), single-electron transfer errors with levels of one part per million were recently measured on a semiconductor SET pump [41].

The ultrastable low-noise current amplifier (ULCA), a new type of transimpedance amplifier developed at PTB, offers unparalleled measurement performance for small direct currents [42, 43]. Among several new versions of the ULCA, designed and tailored for different special applications, a 'low noise' variant with 1 GΩ effective transresistance $A_{TR}$ features 1.4 fA (√Hz)$^{-1}$ current noise and a $1/f$ corner below 1 mHz [44]. The short-term stability of the $A_{TR}$ is about 0.1 µA A$^{-1}$ per day. Provided that $A_{TR}$ is traced to $R_K$ and that the ULCA output voltage $U_{ULCA}$ (100 mV for 100 pA input current sourced by a 'self-referenced' SET pump) is measured by means of a JVS (or a voltmeter traced to $K_J$), a QMT experiment according to $U_{ULCA}(K_J) = A_{TR}(R_K) \times I_{SET}(Q_S)$ can be realized. Different ways in which the electrical quantum standards QHR and JVS can be combined with the ULCA are demonstrated and discussed in [37, 43]. If the calibration of $A_{TR}$ is done on a daily basis, drift of $A_{TR}$ is supressed. Using two of the above-mentioned ULCAs [37] and optimized cryogenic wiring in the refrigerator setup [45], such a setup is capable of closing the QMT with a total uncertainty of 0.1 µA A$^{-1}$ within about 10 h of integration time.

## Acknowledgments

The authors thank S Lotkhov for design and fabrication of the SET devices and G-D Willenberg for the construction of the cryogenic capacitor. Help and support from M Keller (NIST) and B Camarota with the experimental realization and data evaluation as well as for fruitful discussions on the ECCS are gratefully acknowledged. R Behr deserves thanks for support with the voltmeter calibrations and the JVS setup, and V Bürkel, M Busse, and G Muchow for continuous technical support. B Jeanneret, A Eichenberger (both METAS) and X Jehl (CEA Grenoble) kindly supported the work by electronics troubleshooting and supplying spare instrumentation. B Kibble is acknowledged for fruitful discussions on cryocap measurements. D Drung, F Hohls and D Newell (NIST) are acknowledged for helpful discussions on perspectives of QMT experiments.

## Appendix A. Shuttle pumping and hold time measurements on the five-junction *R*-pump

For tuning the SET pump, both needle switches NS1 and NS2 (see figure 1) are opened. In this state, the on-chip SET electrometer monitoring the charge state on the small 'pad' (i.e. the metallic island located between the pump and the electrometer) has maximum resolution and can detect changes on the single-electron scale. The pump is operated in 'shuttle pumping' mode, i.e. cyclically pumping ± one electron onto/from the island, and the feedback voltage $U_{FB}$ corresponding to the electrometer signal is monitored, as shown in figure A1.

Figure A2 shows a typical electrometer signal trace recorded during shuttle pumping at frequency $f_{shuttle}$ = 2.5 MHz (i.e. the repetition frequency of the ac gate voltage pulses applied to the four pump gates A–D, see figure A1) after the dc offsets were tuned. Due to its limited bandwidth of about 1 kHz, the electrometer cannot resolve every single pumping event, but the signal shows single-electron events on the pad caused by relatively rare pumping errors. These single-electron error events can be caused by co-tunnelling, missed cycles, thermal activation, background charge activity or electromagnetic interference in the system [25, 27, 46]. In the electrometer signal trace, pumping errors are indicated by the step-wise jumps, each corresponding to a change of one electron on the charged island. Relating the error rate $\Gamma_{err}$ to the shuttling frequency $f_{shuttle}$ yields the relative transfer error (RTE = $\Gamma_{err}/f_{shuttle}$), i.e. the average error per electron pumped.

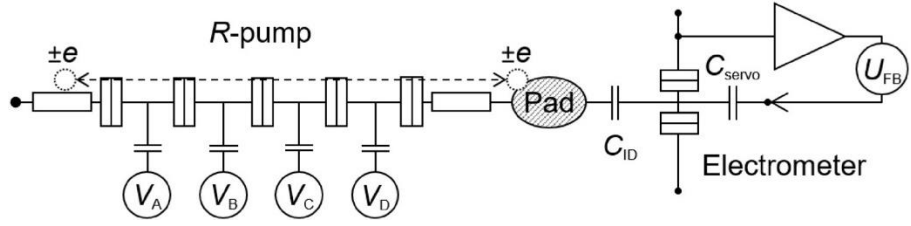

**Figure A1.** Scheme for shuttle-pumping with ± one electron. The *R*-pump pumps one electron in and out from the small metallic 'pad' island between pump and electrometer (see figure 1). The working point of the SET electrometer, capacitively coupled to the pad via an interdigital gate ($C_{ID}$), is locked by a voltage feedback acting on the electrometer servo gate ($C_{servo}$)

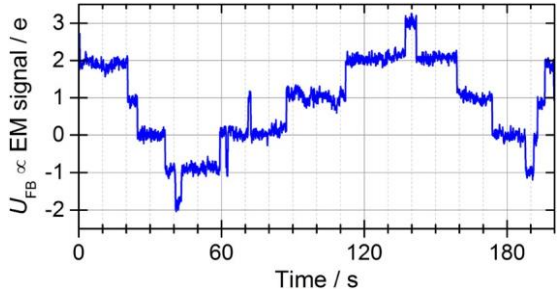

**Figure A2.** Time trace of the electrometer (EM) signal recorded during fast shuttle-pumping ± one electron at $f_{shuttle}$ = 2.5 MHz.
22 ± 5 errors were observed during 200 s, giving an error event rate $\Gamma_{err}$ = (0.11 ± 0.02) s$^{-1}$ with statistical standard uncertainty according to a Poissonian distribution of the error events. Relating the error rate to the shuttling frequency gives the relative transfer error RTE = $\Gamma_{err}/f_{shuttle}$ = (4.4 ± 1) · 10$^{-8}$.

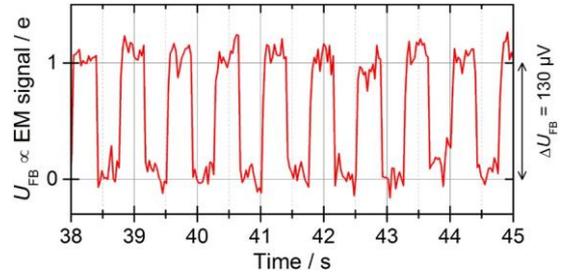

**Figure A3.** Slow shuttle-pumping of ± one electron with a time delay of 0.4 s between each pump event. The changes of the feedback voltage $\Delta U_{FB}$ = 130 μV correspond to the total capacitance of 1.2 fF of the pad island.

Following the pump tuning, a fidelity test of the SET pump is performed in order to verify that the pump properly transfers single electrons. For this, $f_{shuttle}$ is reduced to a few Hz so that the electrometer can follow the single-electron transfer events on the pad. A typical electrometer signal trace recorded during 'slow' shuttle-pumping is shown in figure A3. The electrometer signal clearly follows the periodic charging/d is charging events of single electrons entering and leaving the pad. This signature together with a reasonably low transfer error rate confirms that the pump was properly tuned to high-fidelity pumping performance.

Finally, shuttle-pumping is stopped. In the hold mode, the electrometer detects spontaneous single-electron fluctuations on the pad caused by random tunnelling events through the junction chain of the pump. The dwell times of the electrons on the pad are dependent on background charge activity, temperature or electromagnetic interference in the system. Figure A4 shows a hold time behaviour following to pump tuning, as explained above.

Re-tuning of the *R*-pump was necessary about once per day.

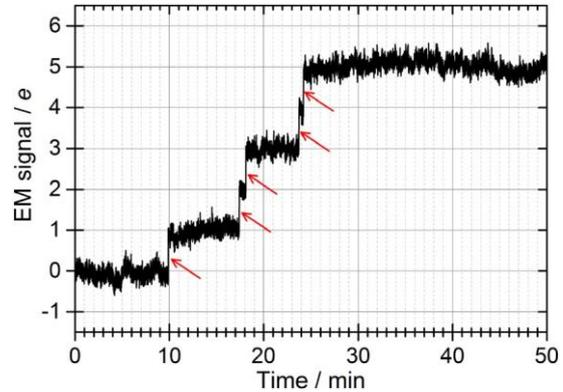

**Figure A4.** Time trace of the electrometer signal with the pump in 'hold' mode. Five single-electron events on the pad (indicated by red arrows) were detected over a duration of 50 min, corresponding to an average hold time of about 428 s. The fact that all tunnel error events in this example are in the same direction indicates a small residual voltage across the pump (see appendix C).

## Appendix B. Quantifying pumping errors

The method for quantifying the pump transfer accuracy explained in appendix A is based on counting all error events occurring in a shuttle-pumping experiment [47], which defines the relative transfer error RTE = $\Gamma_{err}/f_{shuttle}$. Since *every* error event disregarding its direction is accounted, this method is considered the most rigorous way for error quantification. However, this method usually overestimates the effect of pumping errors since it does not take into account the fact that errors in both directions (i.e. extra electrons and deficit electrons on the pad) typically occur, and also that both happen with similar probabilities or rates. As, for example, figure A2 shows, those extra and deficit electron events compensate each other to a certain extent. Physically, the degree of this compensation reflects how well the voltage across the pump was kept close to zero by a voltage feedback circuit during pumping (see figures 1 and A1).

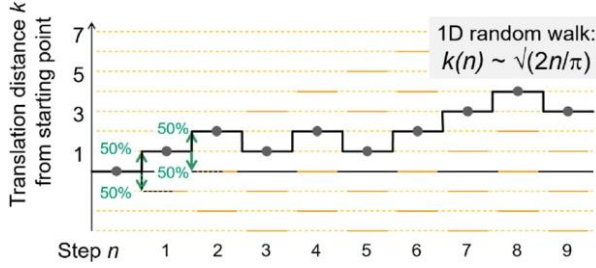

**Figure B1.** Typical state evolution of a 1D trajectory in a random walk scenario. After each step, there is an equal probability for the next step going up or down (Markov process). After $n$ steps, the expectation value $k$ for the distance from the starting point is $\sqrt{(2n/\pi)}$.

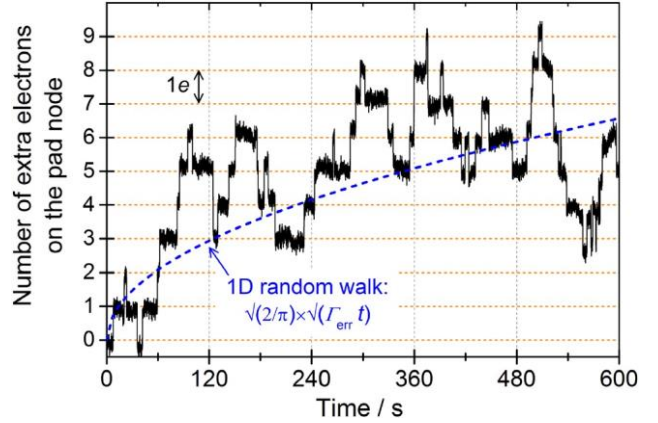

**Figure B2.** Time evolution of the charge on the pad node between SET pump and electrometer during shuttle-pumping at $f_{\text{shuttle}} = 2.5$ MHz. In total $n = 68$ errors occurred over 600 s, so $\Gamma_{\text{err}} = 0.113$ s$^{-1}$. Counting all errors yields the relative transfer error RTE = $\Gamma_{\text{err}}/f_{\text{shuttle}} = 4.5 \times 10^{-8}$. However, during the whole time a net charge difference of only 6 $e$ was accumulated. This agrees with a random-walk-like evolution following $\sqrt{(2n/\pi)} = \sqrt{(2/\pi)} \times \sqrt{(\Gamma_{\text{err}} \times t)}$ (blue dotted line), yielding a net charge change of 6.6 electrons after 600 s and RTE$_{\text{rw}} = 4.4 \times 10^{-9}$.

A more sophisticated approach for error accounting is based on estimating the net error, i.e. the charge accumulated on the pad island during a shuttling experiment [29]. For this, it is assumed that typically the charge state evolution occurring during shuttle-pumping on the pad follows the statistics of a Markov process. This means that after each error event there is an equal probability for the next error going in a positive or a negative direction. Thus, the pad net charge evolution follows a 1D random walk process, as shown in figure B1, and the expectation value $k$ for the translation (distance from the starting point of the trajectory) after $n$ steps is $\sqrt{(2n/\pi)}$ [48].

For shuttle-pumping experiments, the number of steps $n$ is parametrized by the time $t$ according to $n = \Gamma_{\text{err}} t$. With the total number of shuttled electrons $N = f_{\text{shuttle}} t$, the net charge accumulated on the pad (i.e. the net pumping error) is $\sqrt{(2\Gamma_{\text{err}} t/\pi)}$. This defines the quantity RTE$_{\text{rw}}$ for the relative net charge transfer error (the subscript RW stands for 'random walk') according to

$$\text{RTE}_{\text{rw}} = k/N = \sqrt{(2/\pi)} \cdot \sqrt{(\Gamma_{\text{err}}/t)} \; 1/f_{\text{shuttle}} \quad \text{(B.1)}$$

An example is shown in figure B2. Here, transfer error accounting following the 'conservative' approach (counting all errors) yields RTE = $\Gamma_{\text{err}}/f_{\text{shuttle}} = 4.5 \times 10^{-8}$, but the 'random walk' approach gives the significantly lower value RTE$_{\text{rw}} = 4.4 \times 10^{-9}$. For the data shown in figure A2, the evaluation gives RTE = $4.4 \times 10^{-8}$ and RTE$_{\text{rw}} = 7 \times 10^{-9}$, respectively.

For the uncertainty evaluations of the electron-counting phase of the ECCS presented in this paper, however, RTE (not RTE$_{\text{rw}}$) values were used as an estimate for pumping accuracy, giving an upper bound on the relative uncertainty of charge transfer. This conservative approach considers a principle unknown of the experiment: as explained in appendix A, pumping accuracy is determined in a shuttle-pumping experiment preliminary to the capacitor charging phase, with (i) the needle switch NS1 open and (ii) with bidirectional electron pumping. Both conditions, however, are not met during the following cryocap charging phase (see section 3.1.3): then (i) NS1 is closed and (ii) capacitor charging with several millions of electrons is performed by unidirectional pumping.

It cannot be excluded that closing NS1 alters the pumping performance, even if the transfer error measurement is repeated after the capacitor charging sequence as a cross-check.

## Appendix C. Effect of voltage bias across the five-junction $R$-pump

A small voltage $U_{\text{pump}}$ across the SET pump typically causes an asymmetry of the pump errors with respect to the pumping direction during ECCS capacitor charging. In the $U_{\text{cryo}}$ traces, this shows up as asymmetry of $\Delta U_{\text{cryo}}$ for ramping up and down. The effect was investigated by NIST for a seven-junction pump and is discussed in detail in [27]. A quantitative measure $\sigma_{\text{asym}} = \frac{1}{2}(|\langle \Delta U_{\text{pos}}\rangle - \langle \Delta U_{\text{neg}}\rangle|)$ for the asymmetry entering the ECCS uncertainty budget was defined in [11], with

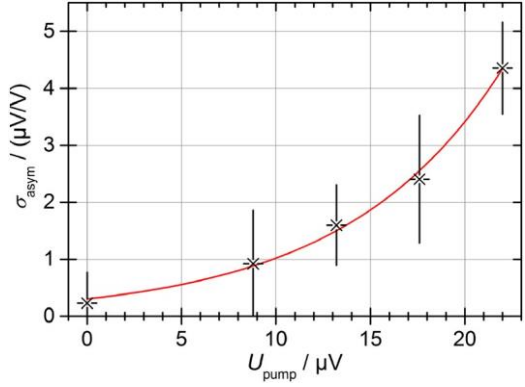

**Figure C1.** Asymmetry $\sigma_{\text{asym}}$ (for definition see text) of the voltage plateaus for up/down ramping in the ECCS capacitor charging phase versus $U_{\text{pump}}$. The red line shows an exponential curve fitted of the data.

$\Delta U_{\text{pos/neg}}$ being the total values of the difference between adjacent voltage plateaus for up/down ramping directions. Analysis of transfer error measurements in shuttle-pumping experiments with the seven-junction pump showed that the transfer errors remained small for $|U_{\text{pump}}| < 20$ μV.

Following the experimental procedure described in [27], the voltage bias effect on the transfer errors of the five-junction $R$-pump was investigated during the capacitor charging phase of an ECCS run (see section 3.1.3). By increasing the voltage $V_{\text{EM}}$ applied to the electrometer servo gate $C_{\text{servo}}$ incrementally (see figure 1), $U_{\text{pump}}$ was changed according to $\Delta U_{\text{pump}} \approx 0.09\, \Delta V_{\text{EM}}$ due to capacitive coupling on the SET chip. The result plotted in figure C1 shows an approximately exponential rise of $\sigma_{\text{asym}}(U_{\text{pump}})$, indicating a corresponding increase of pumping errors with $U_{\text{pump}}$ (see [27]), and $\sigma_{\text{asym}} < 1$ μV V$^{-1}$ for $U_{\text{pump}} < 10$ μV. During the setup of the voltage feedback preliminary to every ECCS run, care was taken that $U_{\text{pump}}$ did not exceed 10 μV (see section 3.1.3).

## Appendix D. Cabling and grounding for cryocap measurements in the bridge phase

For the bridge phase of the ECCS, the 'low potential' center electrode of the coaxial cryocap was connected to the capacitance bridge by closing needle switch NS2, and NS1 is opened to disconnect the SET chip (figure 1). Two coaxial lines, each about 3 m long and composed from different segments, were installed in the dilution refrigerator to connect the cryocap electrodes to the bridge. Each of the lines (including filters) showed a characteristic impedance of 50 Ω, a total (distributed) parallel capacitance of about 310 pF and a total series resistance of about 2 Ω at base temperature. The sequence cable types installed from top to bottom of the refrigerator were as follows:

From the RT terminals of the refrigerator down to the 4 K stage, commercial flexible coax cables with braided shield (type Gore™, about 2 m long) were installed.

For the connections between the 4 K and the mixing chamber stages, commercial superconducting cable segments (type SC-160/50-NbTi–NbTi semi-rigid coaxial cable, each about 40 cm long) were used. Completely superconducting cable was chosen in order to keep the line impedance low and at the same time to suppress heat load on the mixing chamber.

Attenuation of rf interference in both lines was provided by attenuator elements based on modified commercial lowpass filters (Mini-Circuits VLFX-650 three-section low-pass filter, for technical details on the modification see [25]). Per line, two such filter elements were installed in series and thermally anchored at the mixing chamber level, providing a total attenuation of about 60 dB at 1 GHz per line at the base temperature of the refrigerator, and contributing a parallel capacitance of about 100 pF and a series resistance of about 0.9 Ω per line.

The bottom ends of the lines were connected to the cryocap inside the cold box by pieces of semi-rigid coax cables, each about 30 cm long.

Since multiple connections between the outer conductors of the lines can impede capacitance measurement with ac bridges [33], care was taken to avoid any galvanic connection between the shields of the coaxial lines and the body of the cryostat, but to still provide sufficient heat-sinking of the lines at their thermal anchoring points. Technically this was realized by insulating the outer shields of the superconducting coaxial semi-rigid lines against the metal anchors at cryostat potential by using polyimide foil. With this setup, a base temperature of the mixing chamber in the dilution refrigerator of about 30 mK was reached. The whole system comprising cryogenic setup and measurement bridge was connected to the reference potential only via a single grounding point defined by the capacitance bridge.

## Appendix E. Measurements of $C_{\text{cyro}}$ with a commercial multi-frequency bridge

Measurements on the cryocap at base temperature of about 30 mK were performed with a commercial multi-frequency bridge (Andeen-Hagerling model 2700A) at an effective excitation voltage of 15 V(rms). The bridge and the cryocap measurement terminals on top of the refrigerator were connected by coaxial cables (each of 2 m length, with specific serial resistance of 82.5 mΩ m$^{-1}$, serial inductance of 1.4 μH m$^{-1}$, and parallel capacitance of 68.4 pF m$^{-1}$), and the bridge was set to automatically correct the measurement results for the effect caused by these cables. Figure E1 shows the measurement results. The observed non-monotonic frequency dependence with a strong rise above about 10 kHz is attributed to the cryogenic cabling inside the refrigerator, described in appendix D.

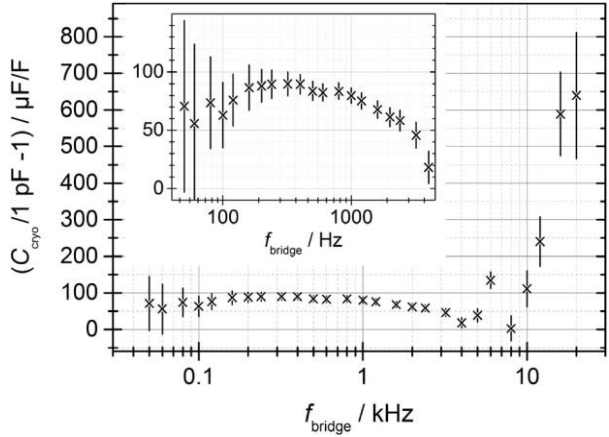

**Figure E1.** Results of $C_{bridge}$ measurements with the commercial multi-frequency bridge. Error bars correspond to type A standard uncertainties (SDOM).

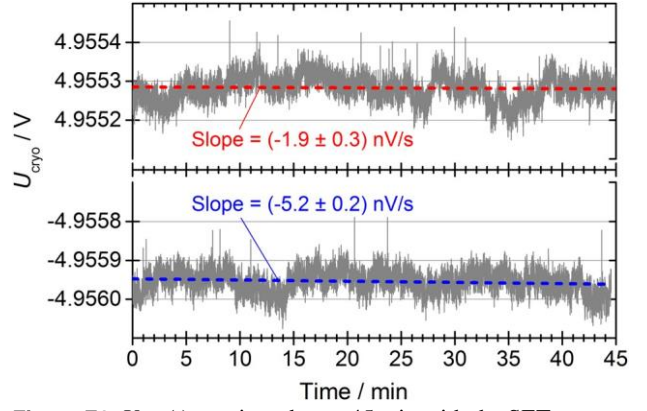

**Figure F1.** $U_{cryo}(t)$ monitored over 45 min with the SET electrometer after the cryocap was charged to ±5 V. The SET pump was in hold mode during these measurements. Dashed lines correspond to linear fits to the data.

Furthermore, measurements were performed with the measurement lines being extended by additional coaxial cable segments, each 3 m in length, in order to simulate the effect of the coaxial lines inside the refrigerator and to quantify their effect. The results showed effects of the order of 10 µF F$^{-1}$ at 1 kHz. Due to the limited accuracy of the multi-frequency bridge, these results, however, could not be used for a high-accuracy quantification with a target uncertainty 1 µF F$^{-1}$ of cable effects in the relevant frequency range below 1 kHz.

## Appendix F. Charge leakage of the cryocap

To determine charge leakage between its electrodes, the cryocap was charged to about ±5 V with the SET pump according to the procedure explained in section 3.1.3. In between, charging was stopped and feedback voltage $U_{cryo}$ was monitored for about 45 min. In the case of charge leakage (for instance between the capacitor electrodes), discharging of the cryocap would be indicated by $U_{cryo}$ moving toward zero for both polarities [11]. Note that the feedback voltage is controlled by the SET electrometer, which measures changes of the pad island potential with respect to ground potential (see figure 1). Therefore, such measurement is only susceptible to charge leakage from the pad node, i.e. the 'low' potential electrode of the cryocap. Possible leakage from the 'high' potential electrode to ground is not detectable and not relevant because it would immediately be compensated by the feedback voltage circuit.

The measurement results presented in figure F1 show small negative slopes for the linear fits to the $U_{cryo}(t)$ traces for both voltage polarities. Also, the observed signal traces are dominated by $1/f$-like fluctuations. Altogether, this indicates that the $U_{cryo}(t)$ behaviour observed was caused by $1/f$ noise and drift of the SET electrometer, confirming earlier findings from the ECCS at PTB presented in [13] and also the NIST results published in [11].

Although charge leakage was below the detectable limit, the data from figure F1 allow estimating an upper bound for leakage effects by considering the difference of the slopes for both polarities $|dU_{cryo}/dt| = 3.3$ nV s$^{-1}$. Attributing this residual drift to a leakage current $I_{leak}$ between the cryocap electrodes yields $I_{leak} = C_{cryo} \times |dU_{cryo}/dt| = 3.3 \times 10^{-21}$ A, corresponding to 0.02 $e$ s$^{-1}$ and a lower limit for the isolation resistance of the cryocap electrodes of $1.5 \times 10^{21}$ Ω. Note that, on average, this means at maximum losing one electron in 50 s, which agrees well with results from pump hold time measurements (see section 3.1.1 and appendix A). Further, considering an ECCS run with ±10 V charging ramps, as shown in figure 3, with about 123.7 million electrons being transferred in about 63 s of ramp time, the corresponding charge loss by leakage is 1.3 electrons, or about 0.01 parts per million in relative units. We therefore assign this value of relative standard uncertainty for charge leakage between the cryocap electrodes in the ECCS uncertainty budget (section 3.4).

## Appendix G. Voltage dependence of $C_{cryo}$

Experimental quantification of the voltage dependence of $C_{cryo}$ was first attempted by single-electron charging experiments following the procedure described in [11]. In subsequent ECCS runs, performed within a short time span of 1 h to minimize the effect from a possible drift of $C_{cryo}$, the cryocap was charged with $N$ electrons corresponding to voltage levels $\Delta U_{cryo}$ of 10 V and 20 V. Values $C_{ECCS} = Ne/\Delta U_{cryo}$ were then calculated as explained in section 3.1.3, and plotted versus $U = Ne/(1\text{ pF})$ in figure G1.

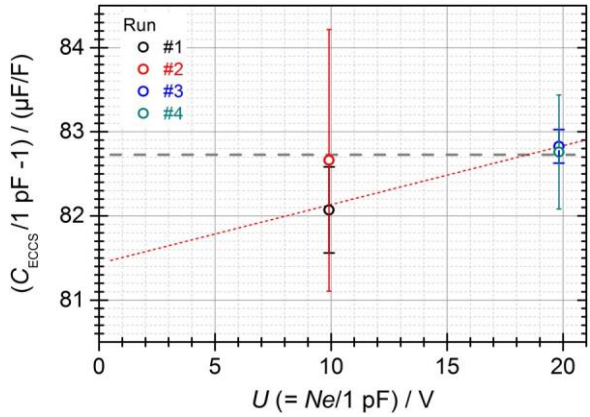

**Figure G1.** $C_{\text{ECCS}}(U)$ resulting from four ECCS runs, plotted as relative difference to 1 pF. The slope of the linear fit (red dotted line) to the data corresponds to a voltage coefficient of $(0.070 \pm 0.014)$ $(\mu\text{F F}^{-1})$ $\text{V}^{-1}$.

The linear fit to the data (red dotted line in figure G1) shows a slope of $(70 \pm 14)$ $(\text{nF F}^{-1})$ $\text{V}^{-1}$. However, deriving a voltage coefficient for $C_{\text{cryo}}$ from this result appeared of limited significance due to the uncertainty of the input data: as the grey horizontal line plotted in the graph indicates, the data are also consistent with zero voltage dependence. Thus, the value of about 70 $(\text{nF F}^{-1})$ $\text{V}^{-1}$ was considered only an estimate of the upper bound on the voltage coefficient of $C_{\text{cryo}}$ (see [8, 11]).

A better estimate for the voltage dependence of the cryocap was later derived from $C_{\text{bridge}}$ measurements with the high-accuracy ratio bridge (see section 3.3.1), taken at excitation voltage levels of 10 V and 20 V (rms). From this, an upper bound of 5 $(\text{nF F}^{-1})$ $\text{V}^{-1}$ was determined for the relative voltage coefficient. The total effective voltages applied to the cryocap span over $\Delta U = 10$ V, with margins given by $\Delta U_{\text{cryo}} = 10$ V and 20 V applied in the ECCS phase (see section 3.1.3) and including $U_{\text{bridge}} = 15$ V (rms) used in the capacitance bridge measurements. Correspondingly, a relative standard uncertainty of 0.05 $\mu\text{F F}^{-1}$ was conservatively attributed for the voltage dependence of $C_{\text{cryo}}$ in the ECCS uncertainty budget (section 3.4).

## Appendix H. Experimental steps to isolate the cryocap as the source of the $C_{\text{cyro}}(f)$ problem

The following list presents the main experimental steps that finally enabled identification of the cryocap as the source of the resonances observed in the high-precision $C_{\text{cyro}}(f)$ measurements, as summarized in section 3.3.1. All measurements listed were performed with the high-accuracy ratio bridge and span over a time period of nearly 2 years.

1. *Question addressed:* **Is there an influence of the 1 K-pot needle valve setting (vibrations)?**

*Diagnostic experiment:* $C_{\text{cryo}}(f)$ was measured at base $T$ (30 mK) in the refrigerator, with 'standard' cryogenic cabling including filters as described in appendix D, and with needle switch NS2 involved. The setting of the 1 K-pot needle valve was varied.

*Result:* Prominent resonance structures in $C_{\text{cryo}}(f)$ with amplitude > 10 $\mu\text{F F}^{-1}$ (see figure 6) independent of needle valve setting.

*Conclusion:* The needle valve setting has no influence on resonances.

2. *Question addressed:* **What is the influence of filtered cryogenic coaxial lines, when cold?**

*Diagnostic experiment:* $C(f)$ of a 1 pF commercial fused silica standard capacitor (Andeen–Hagerling model 1100) was measured at RT. The 'standard' cryogenic cabling including filters (see appendix D) inside the refrigerator was used, cooled down to base temperature. Both lines were connected at the mixing chamber stage, and their RT terminals were used to connect one terminal of the capacitor to the bridge. The other terminal of the capacitor was connected to the bridge with a short coaxial cable. The result from this $C(f)$ measurement was compared to a measurement involving only short coaxial cables at RT. Needle switch NS2 was involved.

*Result:* The filtered cryogenic lines have only a small $C(f)$ effect (< 0.2 $\mu\text{F F}^{-1}$ for 1 kHz < $f$ < 3 kHz), as expected from the cable parameters (cable correction). $C(f)$ showed no resonance structures.

*Conclusion:* The observed $C_{\text{cryo}}(f)$ behaviour is unlikely to be caused by the 'standard' cryogenic lines.

3. *Question addressed:* **What is the influence of needle switch NS2, connecting cryocap and bridge?**

*Diagnostic experiment:* $C_{\text{cryo}}(f)$ was measured at base temperature in the refrigerator with 'standard' cryogenic cabling, but needle switch NS2 was removed from the cabling connecting bridge and cryocap. The cryocap was directly connected to the coaxial cables installed on the refrigerator by a short piece of coaxial cable inside the rf shielding sample box at the mixing chamber stage.

*Result:* Prominent resonance structures in $C_{\text{cryo}}(f)$ (> 13 $\mu\text{F F}^{-1}$ in amplitude) for 1 kHz < $f$ < 3 kHz.

*Conclusion:* The resonance features in $C_{\text{cryo}}(f)$ are not caused by the needle switch.

*4. Question addressed:* **Is there an influence of galvanic connection between cryocap and refrigerator body?**

*Diagnostic experiment:* $C_{\text{cryo}}(f)$ was measured at base temperature in the refrigerator with the *'standard'* cryogenic cabling and needle switch NS2 removed (same as in step 3), but the cryocap was galvanically isolated from the refrigerator body using polyimide foil and plastic mounting screws.

*Result:* Still prominent resonance structures in $C_{\text{cryo}}(f)$ ($> 13$ μF F$^{-1}$ in amplitude) for 1 kHz $< f <$ 3 kHz.

*Conclusion:* A galvanic connection between cryocap and refrigerator body (floating during the $C_{\text{cryo}}$ measurements) has no influence on the resonances in $C_{\text{cryo}}(f)$.

*5. Question addressed:* **What is the frequency dependence of the cryocap when measured at RT, without filtered lines in the refrigerator being involved?**

*Diagnostic experiment:* $C(f)$ of the cryocap was measured with short coaxial cables (all at RT) connecting cryocap and bridge.

*Result:* Small and approximately linear frequency dependency between 1 kHz and 3 kHz without any resonance structures. The measured frequency coefficient of about -70 nF F$^{-1}$ per kHz can be explained by an effect from dielectric layers on the cryocap electrodes.

*Conclusion:* The bare, isolated cryocap shows no $C(f)$ problem when measured at RT.

*6. Question addressed:* **What about the frequency dependence when the cryocap is measured using the filtered lines in the refrigerator, but with the whole system being at RT?**

*Diagnostic experiment:* $C(f)$ of the cryocap was measured in refrigerator setup with *'standard'* cabling (same as in step 3) without NS2, but the whole system was warm (at RT).

*Result:* Negligibly small frequency dependency ($< 0.1$ μF F$^{-1}$) between 1 kHz and 3 kHz, without any resonance structures.

*Conclusion:* The cryocap together with the *'standard'* cabling including filters mounted in the refrigerator setup shows no problem when the whole system is warm (at RT).

*7. Question addressed:* **What about the influence of possible electro-acoustic coupling between 'high'** and 'low' lines in the cabling to the cryocap (see section 3.3.1)?

*Diagnostic experiment:* $C_{\text{cryo}}(f)$ was measured at base temperature in the refrigerator setup with *'standard'* cryogenic cabling including filters, but with outer conductors of semi-rigid cable segments (see appendix D) heavily crimped (about 3 cm distance between adjacent crimping points along the lines). The crimping was done in order to pinch the PTFE insulation, and the strain should avoid *'shaking'* (oscillations) of the inner conductors when the system was cooled and the PTFE was shrinking. Needle switch NS2 was involved.

*Result:* Still prominent resonance structures in $C_{\text{cryo}}(f)$ for 1 kHz $< f <$ 3 kHz.

*Conclusion:* $C_{\text{cryo}}(f)$ resonances are presumably not caused by electroacoustic coupling of the semi-rigid cable segments inside the refrigerator.

*8. Question addressed:* **What if the cryocap is measured at base temperature, but using approved alternative cabling in the refrigerator?**

*Diagnostic experiment:* $C_{\text{cryo}}(f)$ was measured in continuous frequency sweeps (0.5 kHz $< f <$ 2 kHz) at base temperature in the refrigerator setup with new cabling, approved in various other high-accuracy measurements with the ratio bridge. For connecting the cryocap (mounted at the mixing chamber stage) to the RT terminals of the refrigerator, commercial flexible coax cables with braided shield were installed, completely replacing all semi-rigid cable segments of the original *'standard'* cabling (see appendix D). Also, no rf filters were involved in the measurements. Due to the changed wiring, the base temperature of the mixing chamber was slightly elevated by about 10 mK.

*Result:* Prominent resonance structures in $C_{\text{cryo}}(f)$, shown in figure 7.

*Conclusion:* Using approved cryogenic wiring without any filter elements does not remedy the $C_{\text{cryo}}(f)$ problem.

From the experimental results gained in step 3 and, in particular, step 8, it was ruled out that needle switch NS2 caused the $C_{\text{cryo}}(f)$ resonance problem. From the experimental results gained in step 2 and, in particular, step 8, it was ruled out that the cryogenic wiring including the rf filter elements installed inside the refrigerator were the cause of the problem. By exclusion, and in particular from the results of the experiment performed in step 8, it was concluded that the cryocap itself caused the resonances in $C_{\text{cryo}}(f)$. The results from the experiments performed in steps 5 and 6 imply, however, that the problem disappeared when the cryocap was at RT.

## References

[1] Göbel E O and Siegner U 2015 *Quantum Metrology: Foundation of Units and Measurements* (New York: Wiley)

[2] Grabert H and Devoret M H (ed) 1992 *Single Charge Tunneling—Coulomb Blockade Phenomena in Nanostructures* (*NATO ASI Series B*) vol 294 (New York: Plenum)

[3] Pekola J P, Saira O-P, Maisi V F, Kemppinen A, Möttönen M, Pashkin Y A and Averin D V 2013 Single-electron current sources: toward a refined definition of the ampere *Rev. Mod. Phys.* **85** 1421–72

[4] Piquemal F and Genevès G 2000 Argument for a direct realization of the quantum metrological triangle *Metrologia* **37** 207–11

[5] Resolution 1 of the 25th CGPM 2014 www.bipm.org/en/CGPM/db/25/1/
Draft of the ninth SI Brochure 2015 www.bipm.org/utils/common/pdf/si-brochure-draft-2016.pdf

[6] Likharev K K and Zorin A B 1985 Theory of the Bloch-wave AQ7 oscillations in small Josephson junctions *J. Low Temp. Phys.* **59** 347–82

[7] Williams E R, Ghosh R N and Martinis J M 1992 Measuring the electron's charge and the fine-structure constant by counting electrons on a capacitor *J. Res. Natl Inst. Stand. Technol.* **97** 299–304

[8] Keller M, Eichenberger A, Martinis J and Zimmerman N 1999 A capacitance standard based on counting electrons *Science* **285** 1706–9

[9] Scherer H, Lotkhov S V, Willenberg G-D and Zorin A B 2005 Steps toward a capacitance standard based on singleelectron counting at PTB *IEEE Trans. Instrum. Meas.* **54** 666–9

[10] Hof C, Jeanneret B, Eichenberger A, Overney F, Keller M W and Dalberth M J 2005 Manipulating single electrons with a seven junction pump *IEEE Trans. Instrum. Meas.* **54** 670–2

[11] Keller M W, Zimmerman N M and Eichenberger A L 2007 Uncertainty budget for the NIST electron counting capacitance standard, ECCS-1 *Metrologia* **44** 505–12

[12] Kemppinen A *et al* 2010 Development of the SINIS turnstile for the quantum metrological triangle *2010 Conf. on Precision Electromagnetic Measurements Digest* pp 125–6

[13] Camarota B, Scherer H, Keller M W, Lotkhov S V, Willenberg G-D and Ahlers F J 2012 Electron counting capacitance standard with an improved five-junction R-pump *Metrologia* **49** 8–14

[14] Devoille L, Feltin N, Steck B, Chenaud B, Sassine S, Djordevic S, Séron O and Piquemal F 2012 Quantum metrological triangle experiment at LNE: measurements on a three-junction R-pump using a 20 000:1 winding ratio cryogenic current comparator *Meas. Sci. Technol.* **23** 124011

[15] Scherer H and Camarota B 2012 Quantum metrology triangle experiments: a status review *Meas. Sci. Technol.* **23** 124010
Scherer H and Camarota B 2013 A review on quantum metrology triangle experiments *IEEE/ CSC ESAS Superconductivity NEWS FORUM (global Edition)* http://snf.ieeecsc.org/abstracts/cr31-review-quantum-metrology-triangle-experiments

[16] Scherer H, Lotkhov S V, Willenberg G-D and Camarota B 2009 Progress towards the electron counting capacitance standard at PTB *IEEE Trans. Instrum. Meas.* **58** 997–1002

[17] Lotkhov S V, Camarota B, Scherer H, Weimann T, Hinze P and Zorin A B 2009 Shunt-protected singleelectron tunneling circuits fabricated on a quartz wafer *Nanotechnology Materials and Devices Conf. (NMDC '09)* (IEEE) pp 23–6

[18] Camarota B, Lotkhov S V, Scherer H, Weimann T, Hinze P and Zorin A B 2010 Properties of shunt-protected tunneling devices for the electron counting capacitance standard (ECCS) experiment at PTB *2010 Conf. on Precision Electromagnetic Measurements Digest* pp 291–2

[19] Lotkhov S V, Bogoslovsky S A, Zorin A B and Niemeyer J 2001 Operation of a three-junction single-electron pump with on-chip resistors *Appl. Phys. Lett.* **78** 946–8

[20] Willenberg G-D and Warnecke P 2001 Stable cryogenic vacuum capacitor for single-electron charging experiments *IEEE Trans. Instrum. Meas.* **50** 235–7

[21] Zimmerman N M 1996 Capacitors with very low loss: cryogenic vacuum-gap capacitors *IEEE Trans. Instrum. Meas.* **45** 841–46

[22] Zimmerman N M, Simonds B J and Wang Y 2006 An upper bound to the frequency dependence of the cryogenic vacuum-gap capacitor *Metrologia* **43** 383–8

[23] Mohr P J, Taylor B N and Newell D B 2012 CODATA recommended values of the fundamental physical constants: 2010 *Rev. Mod. Phys.* **84** 1527–605

[24] Mohr P, Taylor B and Newell D 2016 CODATA recommended values of the fundamental physical constants: 2014 *Rev. Mod. Phys.* **88** 35009

[25] Keller M W 2009 Practical aspects of counting electrons with a single-electron tunneling pump *Eur. Phys. J. Spec. Top.* **172** 297–309

[26] Keller M W, Martinis J M, Steinbach A H and Zimmerman N M 1997 A seven-junction electron pump: design, fabrication, and operation *IEEE Trans. Instrum. Meas.* **46** 307–10

[27] Jehl X, Keller M, Kautz R, Aumentado J and Martinis J 2003 Counting errors in a voltage-biased electron pump *Phys. Rev.* B **67** 165331

[28] Keller M W, Martinis J M and Zimmerman N M 1996 Accuracy of electron counting using a 7-junction electron pump *Appl. Phys. Lett.* **69** 1804–6

[29] Scherer H, Camarota B, Keller M W and Lotkhov S V 2012 Improved performance of the ECCS

experiment at PTB *2012 Conf. on Precision Electromagnetic Measurements Digest* pp 350–1

[30] Keller M, Martinis J and Kautz R 1998 Rare errors in a wellcharacterized electron pump: comparison of experiment and theory *Phys. Rev. Lett.* **80** 4530–3

[31] Kemppinen A, Lotkhov S V, Saira O P, Zorin A B, Pekola J P and Manninen A J 2011 Long hold times in a two-junction electron trap *Appl. Phys. Lett.* **99** 1–4

[32] Behr R, Palafox L, Ramm G, Moser H and Melcher J 2007 Direct comparison of Josephson waveforms using an ac quantum voltmeter *IEEE Trans. Instrum. Meas.* **56** 235–8

[33] Awan S, Kibble B and Schurr J 2011 *Coaxial Electrical Circuits for Interference-Free Measurements* (London: IET electrical measurement series/Institution of Engineering and Technology)

[34] Schurr J, Bürkel V and Kibble B P 2009 Realizing the farad from two ac quantum Hall resistances *Metrologia* **46** 619–28

[35] Keller M W 2008 Current status of the quantum metrology triangle *Metrologia* **45** 102–9

[36] Stein F *et al* 2015 Validation of a quantized-current source with 0.2 ppm uncertainty *Appl. Phys. Lett.* **107** 103501

[37] Stein F *et al* 2017 Robustness of single-electron pumps at sub-ppm current accuracy level *Metrologia* **54** S1–8

[38] Fricke L *et al* 2013 Counting statistics for electron capture in a dynamic quantum dot *Phys. Rev. Lett.* **110** 126803

[39] Fricke L *et al* 2014 Self-referenced single-electron quantized current source *Phys. Rev. Lett.* **112** 226803

[40] Wulf M 2013 Error accounting algorithm for electron counting experiments *Phys. Rev.* B **87** 35312

[41] Giblin S P, See P, Petrie A, Janssen T J B M, Farrer I, Griffiths J P, Jones G A C, Ritchie D A and Kataoka M 2016 High-resolution error detection in the capture process of a single-electron pump *Appl. Phys. Lett.* **108** 23502

[42] Drung D, Krause C, Becker U, Scherer H and Ahlers F J 2015 Ultrastable low-noise current amplifier: a novel device for measuring small electric currents with high accuracy *Rev. Sci. Instrum.* **86** 24703

[43] Drung D, Götz M, Pesel E and Scherer H 2015 Improving the traceable measurement and generation of small direct currents *IEEE Trans. Instrum. Meas.* **64** 3021–30

[44] Drung D and Krause C 2016 Ultrastable low-noise current amplifiers with extended range and improved accuracy *IEEE Trans. Instrum. Meas.* http://ieeexplore.ieee.org/stamp/stamp.jsp?arnumber=7580617

[45] Krause C, Scherer H and Drung D 2016 Cable noise investigations for high-accuracy measurements of small direct currents *2016 Conf. on Precision Electromagnetic Measurements* pp 210–1
Krause C, Drung D and Scherer H 2017 Measurement of sub-picoampere direct currents with uncertainties below ten attoamperes *Rev. Sci. Instrum.* **88** 24711

[46] Kautz R L, Keller M W and Martinis J M 2000 Noise-induced leakage and counting errors in the electron pump *Phys. Rev.* B **62** 15888–902

[47] Kautz R L, Keller M W and Martinis J M 1999 Leakage and counting errors in a seven-junction electron pump *Phys. Rev* B **60** 8199–212

[48] Pearson K 1905 The problem of the random walk *Nature* **72** 294–94
van Kampen N G 1992 *Stochastic Processes in Physics and Chemistry* Revised and enlarged edn (Amsterdam: North-Holland)